\documentclass[aps,prl,twocolumn,superscriptaddress,longbibliography]{revtex4-1}
\usepackage{graphicx}
\usepackage{latexsym}
\usepackage{amssymb}
\usepackage{amsmath}
\usepackage{amsfonts}
\usepackage{upgreek}
\usepackage{float}
\usepackage{bm}
\usepackage{multirow}
\usepackage{color}
\usepackage[T1]{fontenc}
\usepackage{hyperref}
\usepackage{subfigure}

\hypersetup{
colorlinks = true,
linkcolor = [rgb]{0.70,0.13,0.13},
citecolor = [rgb]{0.13,0.55,0.13},
urlcolor  = [rgb]{0.25, 0.41, 0.88}}

\begin{document}

\title{Dynamical scaling of Loschmidt echo in non-Hermitian systems}
\author{Jia-Chen Tang}
\affiliation{College of Science, Nanjing University of Aeronautics and Astronautics, Nanjing, 211106, China}
\affiliation{Key Laboratory of Aerospace Information Materials and Physics (Nanjing University of Aeronautics and Astronautics), MIIT, Nanjing 211106, China}

\author{Su-Peng Kou}
\affiliation{Center for Advanced Quantum Studies, Department of Physics, Beijing Normal University, Beijing 100875, China}

\author{Gaoyong Sun}
\thanks{Corresponding author: gysun@nuaa.edu.cn}
\affiliation{College of Science, Nanjing University of Aeronautics and Astronautics, Nanjing, 211106, China}
\affiliation{Key Laboratory of Aerospace Information Materials and Physics (Nanjing University of Aeronautics and Astronautics), MIIT, Nanjing 211106, China}

\begin{abstract}
We show that non-Hermitian biorthogonal many-body phase transitions can be characterized by the enhanced decay of Loschmidt echo.
The quantum criticality is numerically investigated in a non-Hermitian transverse field Ising model by performing the finite-size dynamical scaling of Loschmidt echo.
We determine the equilibrium correlation length critical exponents that are consistent with previous results from the exact diagonalization. 
More importantly, we introduce a simple method to detect quantum phase transitions with the short-time average of rate function motivated by the critically enhanced decay behavior of Loschmidt echo. 
Our studies show how to detect equilibrium many-body phase transitions with biorthogonal Loschmidt echo that can be observed in future experiments via quantum dynamics after a quench.

\end{abstract}

\maketitle

%%%%%%%%%%%%%%%%%%%%%%%%%%%%%%%
% Introduction
%\section{Introduction}

\textit{Introduction}.- The research on quantum phase transitions is one of the central issues in condensed matter physics \cite{sachdev1999quantum}. 
One goal of current research is to discover novel quantum matters and quantum phase transitions \cite{levin2006detecting}.
Recently, an active research field is to investigate interesting quantum phases and quantum phase transitions in non-Hermitian systems.
Previous studies show that non-Hermitian systems can exhibit a lot of fascinating phenomena without counterpart in Hermitian systems \cite{bergholtz2021exceptional,ashida2021non},
such as the bulk-boundary correspondence breakdown and the non-Hermitian skin effect \cite{lee2016anomalous,yao2018edge,kunst2018biorthogonal,xiong2018does,
gong2018topological,alvarez2018non,yokomizo2019non,okuma2020topological,zhang2020correspondence,yang2020non,
wang2020defective,jiang2020topological,weidemann2020topological,xiao2020non,borgnia2020non}, 
exceptional points and the bulk Fermi arcs \cite{heiss2012physics,kozii2017non,hodaei2017enhanced,zhou2018observation,miri2019exceptional,park2019observation,yang2019non,
ozdemir2019parity,dora2019kibble,jin2020hybrid,xiao2021observation}.
The extension to non-Hermitian interacting many-body systems are also explored to understand the effect of non-Hermiticity \cite{matsumoto2020continuous,yang2020anomalous,ashida2017parity,chang2019entanglement,lee2020many,pan2020non,pan2020interaction,xu2020topological,
zhang2020skin,lee2021many,shackleton2020protection,liu2020non,yang2021exceptional,herviou2019entanglement,yoshida2019non,yoshida2020fate}.
It is shown that new types of phase transitions can occur between gapped phases without gap closing in non-Hermitian many-body models \cite{matsumoto2020continuous,yang2020anomalous}.
Consequently, a key issue is to find the phase transition and figure out the nature of the quantum criticality. 

In Hermitian systems, a second-order phase transition can be described by a phenomenological order parameter according to the Landau-Ginzburg theory.
Hence, a Hermitian system usually undergoes a phase transition with the gap closing from an ordered phase with a nonzero order parameter to a disordered phase 
with a vanishing order parameter. 
Thanks to the development of quantum information science, quantum phase transitions and critical phenomena can also be understood with the concepts from quantum information, 
i.e. the quantum entanglement \cite{osterloh2002scaling,horodecki2009quantum,eisert2010colloquium}, 
the quantum fidelity \cite{Zanardi2006,You2007,Venuti2007,Gu2010,Albuquerque2010,Sun2017,Zhu2018,Wei2018,Chen2008,Gu2008,Yang2008,Lu2018,Sun2015} 
and the Loschmidt echo (LE) \cite{Quan2006decay,Hwang2019Universality,mukherjee2012loschmidt,karl2017universal,pelissetto2018dynamic,nigro2019scaling,titum2019probing,halimeh2021local,daug2021dynamical,rossini2021coherent}.
A natural question is whether such approaches can also be used to characterize the non-Hermitian many-body phase transitions.

Recently, above approaches were extended to non-Hermitian systems to identify phase transitions \cite{chang2019entanglement,herviou2019entanglement,sun2022biorthogonal,tzeng2021hunting,solnyshkov2021quantum,zhang2021quantum,para2021probing,xu2021dynamical,
zhou2018dynamical,zhou2021non,zhai2020out}. 
Here, we are interested in non-Hermitian many-body systems with real eigenvalues whereby we can define the ground state,
and focus on the LE that can be realized in experiments via a quench dynamics \cite{xiao2020non,qiu2019fixed,wang2019observation}.
The LE is defined as the overlap between an initial ground-state and its time-evolved state, which exhibits a decay and revival behavior influenced by the quantum criticality \cite{Quan2006decay,Hwang2019Universality}. 

%%% We show that the LE in biorthogonal bases can serve as a probe of detecting phase transitions of non-Hermitian systems because of above features. 

In this paper, we study the dynamical scaling laws of biorthogonal LE in close proximity to the critical point of one-dimensional non-Hermitian transverse field Ising model.
We perform the finite-size scaling theory and demonstrate that the system undergoes a second-order phase transition with the Ising universal class by numerically determining
the equilibrium correlation length critical exponents of the model.
What is more, we introduce the time average of rate function to illustrate how to study the quantum criticality without knowing the critical value 
or even without assuming the phase transition existence in advance.
Consequently, the biorthogonal LE can serve as a simple probe of discovering the non-Hermitian many-body phase transitions due to the critically enhanced decay behavior.

%This paper is organized as follows. 
%In Sec.\ref{sec:BLE}, we introduce the concept and the dynamical scaling law of biorthogonal LE. 
%In Sec.\ref{sec:NHTI}, we investigate the non-Hermitian transverse field Ising chain to verify the scaling law and the critically enhanced decay of the biorthogonal LE. 
%In Sec.\ref{sec:Res}, we present the results.
%In Sec.\ref{sec:Con}, we summarize.

%%%%%%%%%%%%%%%%%%%%%%%%%%%%%%%
% Loschmidt echo
%\section{Biorthogonal Loschmidt echo}
%\label{sec:BLE}
\textit{Biorthogonal Loschmidt echo}.- Given a general non-Hermitian system that is described by the Hamiltonian, 
\begin{align}
H(\lambda) = H_{0}+ \lambda H_{1},
\end{align}
with a control parameter $\lambda$, and the $H(\lambda) \neq H^{\dagger}(\lambda)$.
The time-independent Schr\"{o}dinger equations for the $H(\lambda)$ and $H^{\dag}(\lambda)$ can be written as, \cite{sun2022biorthogonal,brody2013biorthogonal,sternheim1972non}
\begin{align}
H(\lambda) | \psi_{j}^{R}(\lambda) \rangle = E_{j}(\lambda) | \psi_{j}^{R}(\lambda) \rangle,  \label{eigR} \\
H^{\dag}(\lambda) | \psi_{j}^{L}(\lambda) \rangle = E_{j}^{\ast}(\lambda) | \psi_{j}^{L}(\lambda) \rangle,
\label{eigL}
\end{align}
where the $E_{j}(\lambda)$, $E_{j}^{\ast}(\lambda)$ and the $| \psi_{j}^{R}(\lambda) \rangle $, $ | \psi_{j}^{L}(\lambda) \rangle$
are eigenvalues and eigenvectors of the right and left eigenvectors of the Hamiltonian $H(\lambda)$ and $H^{\dagger}(\lambda)$ respectively.
And the eigenvectors obey the bi-orthonormal relation and completeness relation, \cite{sun2022biorthogonal,brody2013biorthogonal,sternheim1972non}
\begin{align}
\langle \psi_{i}^{L}(\lambda)| \psi_{j}^{R}(\lambda) \rangle = \delta_{ij},  \\
\sum_{j} | \psi_{j}^{R}(\lambda) \rangle \langle \psi_{j}^{L}(\lambda)| = 1.
\end{align}
The time-evolved states $| \psi_{j}^{R}(\lambda_f,\lambda_i, t) \rangle$ and $| \psi_{j}^{L}(\lambda_f,\lambda_i, t) \rangle$ after a quench from $\lambda_i$ to $\lambda_f$ are obtained as,
\begin{align}
| \psi_{j}^{R}(\lambda_f,\lambda_i, t) \rangle = e^{-iH(\lambda_f)t}| \psi_{j}^{R}(\lambda_i) \rangle, \label{EvoR} \\
| \psi_{j}^{L}(\lambda_f,\lambda_i, t) \rangle = e^{-iH^{\dagger}(\lambda_f)t}| \psi_{j}^{L}(\lambda_i) \rangle,
\label{EvoL}
\end{align}
by evolving the initial right eigenstates $| \psi_{j}^{R}(\lambda_i) \rangle$ and left eigenstates $|\psi_{j}^{L}(\lambda_i) \rangle $ from time $t=0$.

In the following, we will focus on the time evolution of ground states $| \psi_{0}^{R}(\lambda_i) \rangle$ and $| \psi_{0}^{L}(\lambda_i) \rangle$ and
introduce the biorthogonal LE as,
\begin{align}
L(\lambda_f,\lambda_i,t)=\langle \psi_{0}^{L}(\lambda_i)|\psi_{0}^{R}(\lambda_f,\lambda_i,t) \rangle \langle \psi_{0}^{L}(\lambda_f,\lambda_i,t)|\psi_{0}^{R}(\lambda_i) \rangle
\label{BioLE}
\end{align}
where the bi-orthonormal relation $\langle \psi_{0}^{L}(\lambda_i)| \psi_{0}^{R}(\lambda_i) \rangle = 1$ has been imposed. 
If the Hamiltonian is Hermitian, $H(\lambda) = H^{\dagger}(\lambda)$, we get the usual LE, $L(\lambda_f,\lambda_i,t)=|\langle \psi_{0}(\lambda_i)|e^{-iH(\lambda_f)t}|\psi_{0}(\lambda_i) \rangle|^2$.
In Hermitian systems, the decay of LE is enhanced by quantum criticality \cite{Quan2006decay}.
In addition, the LE of Hermitian systems in close proximity to the critical point $\lambda_c$ is scaling invariance \cite{Hwang2019Universality} ,
\begin{align}
L(\widetilde{N},\widetilde{\delta\lambda},\widetilde{g},\widetilde{t}) = L(N,\delta\lambda, g, t),
\label{LEscale}
\end{align}
under the scaling transformation,
\begin{align}
\widetilde{N}= b^{-1}N, ~~ \widetilde{\delta\lambda}= b^{1/\nu}\delta\lambda, ~~ \widetilde{g}= b^{1/\nu}g, ~~ \widetilde{t}= b^{-z}t, 
\label{ParaTrans}
\end{align}
from the renormalization group analysis.  Here, the $1/b$ denotes the lattice sites grouped into a block, and the $N$ is the lattice size.
The $\nu$ and $z$ are the correlation length critical exponent and the dynamical critical exponent, respectively.
The $\delta\lambda$ and $g$ defined as,
\begin{align} 
\delta\lambda=\lambda_c-\lambda_i,  \\
g=\lambda_i -\lambda_f,
\end{align}
are control parameters for quench dynamics.
We show next that the scaling invariance law and the enhanced decay of biorthogonal LE persist in non-Hermitian many-body systems.

%%%%%%%%%%%%%%%%%%%%%%%%%%%%%%%
% Model
%\section {Model}
%\label{sec:NHTI}
\textit{Model}.- To verify the scaling invariance and the critically enhanced decay of biorthogonal LE, we consider an one dimensional non-Hermitian ferromagnetic transversed field Ising (NHTI) chain defined as,
\begin{align}
H = {}& -\sum_{j=1}^{N} J \sigma_{j}^{x} \sigma_{j+1}^{x} +  \sum_{j=1}^{N} \lambda (\sigma_{j}^{z} + i \gamma \sigma_{j}^{y}),
\label{HamEff}
\end{align}
where $J>0$, $\lambda > 0$, $\gamma \geq 0$ are control parameters, and $i= \sqrt{-1}$ denotes the imaginary unit.
Here, $\sigma_{j}^{x}, \sigma_{j}^{y}, \sigma_{j}^{z}$ are three Pauli matrices at the $j$th site along $x,y,z$ directions, and $N$ is the lattice size.
For $\gamma = 0$, the system is a conventional transversed field Ising model with a second-order phase transition at $\lambda/J = 1$ between the ferromagnetic phase and the paramagnetic phase. 
However, for $\gamma \neq 0$, the system becomes non-Hermitian and has an exceptional point at $\gamma_c = 1$ separated by a parity-time (PT) symmetry ($\gamma < 1$) regime
and a broken PT symmetry ($\gamma > 1$) regime \cite{yang2020anomalous,zhang2020topological} because of the imaginary transverse field term $i \gamma \sigma_{j}^{y}$.
The magnetic field $i \gamma \sigma_{j}^y$ describes the gain from (or loss to) the environment, which can be understood as an effective non-Hermitian Hamiltonian 
of the Lindblad equation \cite{bergholtz2021exceptional,ashida2021non} and realized by the optical pumping in a three level system \cite{lee2014heralded}.
 
In the PT symmetry ($\gamma < 1$) regime, all eigenvalues of the model are real.
In addition, it is shown that a phase transition occurs at the gap closing point,
\begin{align}
\lambda_c = \sqrt{1/(1-\gamma^2)}
\end{align}
between the biorthogonal ferromagnetic phase 
and the biorthogonal paramagnetic phase \cite{yang2020anomalous,zhang2020topological}.
In the following, we will focus on this PT symmetry ($\gamma < 1$) regime so that we can define the ground state eigenvector and the biorthogonal LE via the energy minimum.

\begin{figure}%[ht]
\includegraphics[width=8.6cm]{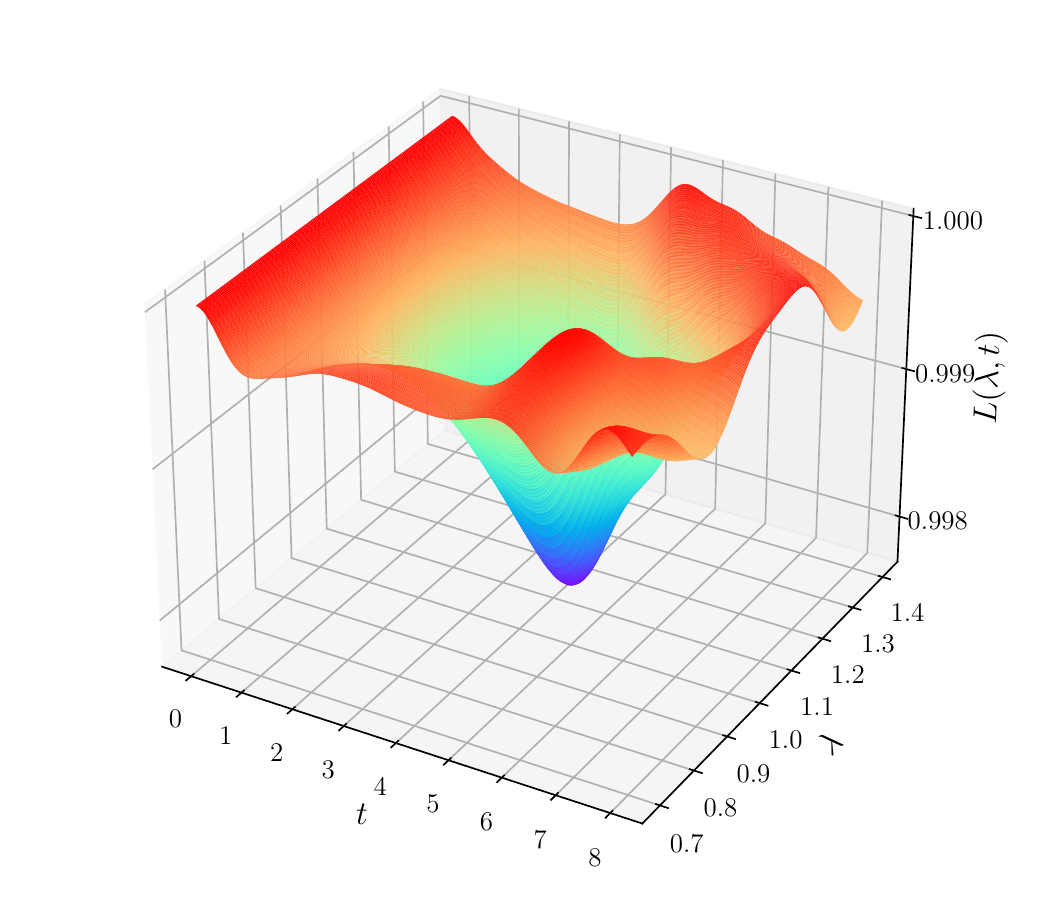}
\caption{(color online) Time evolution of the biorthogonal LE $L(\lambda,t)$ as the function of $\lambda$ and $t$ for $N=16$ lattice sites with $\gamma=0.5$.
It indicates that the decay of biorthogonal LE is enhanced by the quantum criticality and is consistent with the analytical result.}
\label{LEfig}
\end{figure}

\begin{figure}%[ht]
\includegraphics[width=8.6cm]{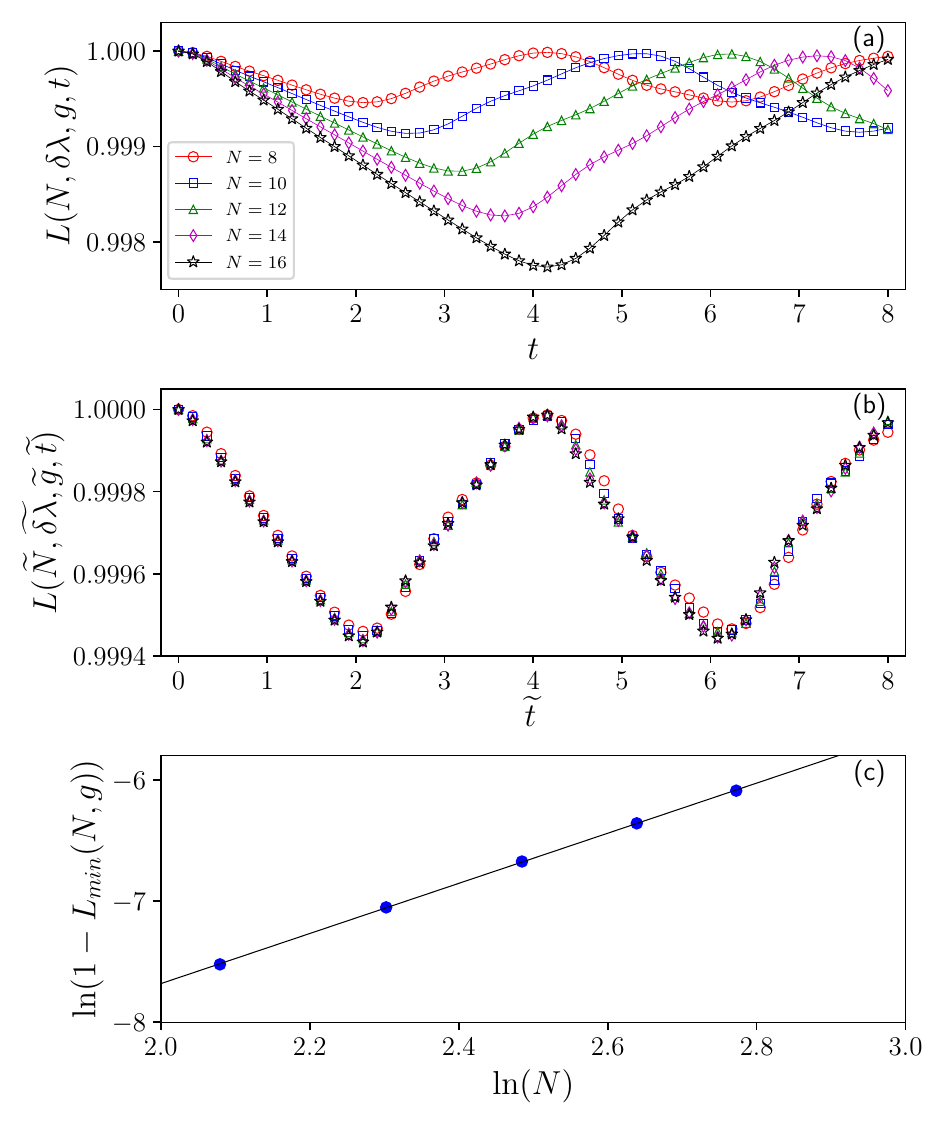}
\caption{ (Color online) Dynamical scaling of the biorthogonal LE at $\gamma=0.5$ in the NHTI chain. 
(a) The biorthogonal LE $L(N,\delta \lambda,g,t)$ with $\delta \lambda=0.02$ and $g=0.01$ as a function of time 
for different system sizes, $N =8,10,12,14, 16$ (from bottom to top). 
(b) Scaling invariance of the biorthogonal LE $L(\widetilde{N},\widetilde{\delta \lambda},\widetilde{g},\widetilde{t})$ given in Eq.(\ref{LEscale}). 
We take $N=8$, $\delta\lambda=0.02,g=0.01$, $\nu=z=1$ and $b=1$. All other data for $b=4/5,2/3,4/7,1/2$ collapse onto a single curve. 
(c) Finite-size scaling of the $1-L_{min}$ obtained from (a) as a function of lattice size $N$. 
The blue circle symbols are the numerical results, and the black solid line is the fitting curve.
The critical exponent of the correlation length derived from fitting curve is $\nu=0.968$.}
\label{collapsefig}
\end{figure}

%%%%%%%%%%%%%%%%%%%%%%%%%%%%%%%
% Results
%\section {Results}
%\label{sec:Res}
\textit{Results}.- To compute the ground state and the biorthogonal LE of the NHTI model, we perform the exact diagonalization that can be generated to arbitrary non-integrable models \cite{wang2020effective,gopalakrishnan2021entanglement} with periodic boundary conditions $\sigma_{N+1}^{x} =\sigma_{1}^{x}$.
Without loss of generality, we choose $J=1$ and $\gamma=0.5$ for simplicity in our numerical simulations. 
The ground states are found separately from Eq.(\ref{eigR}) and Eq.(\ref{eigL}) by the energy minimum as Hermitian models.
The time evolution of the right and left ground states are obtained from Eq.(\ref{EvoR}) and Eq.(\ref{EvoL}) independently.
To see the decay of the biorthogonal LE, we quench the system from an initial $\lambda_i$ to a final $\lambda_f$ with a small constant step $g=-0.01$ and $\Delta t=0.02$.
We note that the $\Delta t$ scales as $\Delta \widetilde{t}= b^{-z} \Delta t$ from the scaling law in Eq.(\ref{ParaTrans}) during the simulations.
The corresponding biorthogonal LE are calculated from Eq.(\ref{BioLE}) by varying the initial $\lambda_i$. 
The data of the biorthogonal LE are presented in Fig.\ref{LEfig}, where a deep valley appears around the critical point $\lambda_c = 2/\sqrt{3} \approx 1.155$ during the time evolution.
This implies that the decay of the biorthogonal LE is highly enhanced by the quantum criticality indicating that the biorthogonal LE can in principle characterize 
the biorthogonal many-body phase transitions.

\begin{figure}%[ht]
\includegraphics[width=8.6cm]{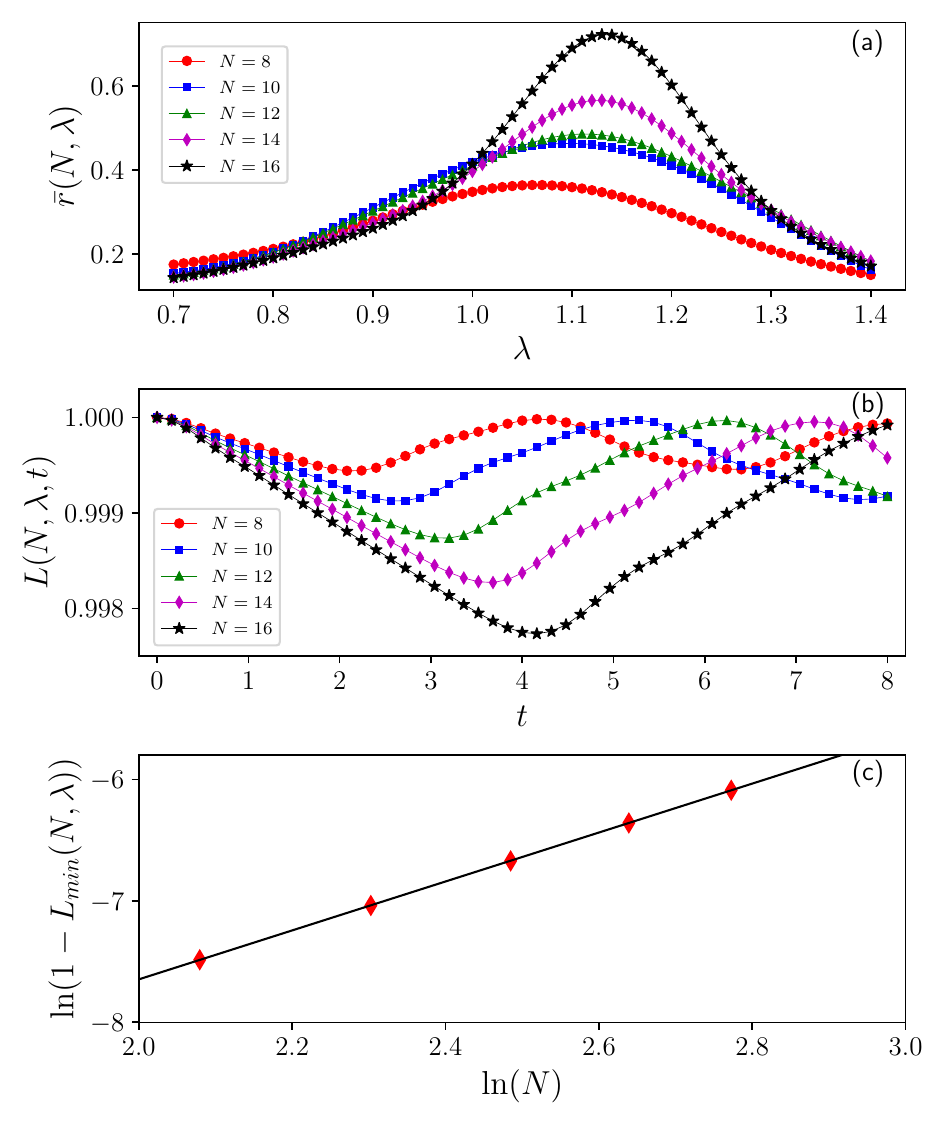}
\caption{ (Color online) Scaling of the rate function and biorthogonal LE in the NHTI model with $\gamma=0.5$. 
(a) The short-time average rate function $\bar{r}(N, \lambda)$ with respect to $\lambda$ for different lattice sizes, $N =8,10,12,14, 16$ (from bottom to top). 
(b) The biorthogonal LE $L(N,\lambda,t)$ at the peak position of $\bar{r}(N, \lambda)$ in (a) with $g=-0.01$ as a function of time 
for different lattice sizes, $N =8,10,12,14, 16$ (from top to bottom). 
(c) Finite-size scaling of the $1-L_{min}$ obtained from (b) as a function of lattice size $N$. 
The red diamond symbols are the numerical values, while the black solid line denotes the fitting curve.
The correlation length critical exponent obtained from fitting curve is $\nu=0.993$. }
\label{ratefig}
\end{figure}

From the renormalization group analysis, the biorthogonal LE is scaling invariance near the critical point under the transformation Eq.(\ref{ParaTrans}). 
To verify the scaling invariance Eq.(\ref{LEscale}), we first perform numerical calculations at $\lambda = 1.15$ (exact $\lambda_c \approx 1.155$)
by choosing the parameters $\delta\lambda=0.02$, $g=0.01$ and $\gamma=0.5$ for $N=8,10,12,14,16$. 
The data are represented in Fig.\ref{collapsefig}(a), where the biorthogonal LEs exhibit a decay and revival dynamics separately as Hermitian systems \cite{Quan2006decay,Hwang2019Universality}. 
However, if the biorthogonal LEs are derived using the transformation Eq.(\ref{ParaTrans}) instead, 
all LEs collapse onto a single curve using the Ising universal class $\nu=z=1$ as shown in Fig.\ref{collapsefig}(b), confirming the validity of scaling invariance in Eq.(\ref{LEscale}).

It was shown in Ref.[\onlinecite{Hwang2019Universality}], under the conditions $\delta\lambda=0$ and $N^{1/\nu}g \ll 1$, the minimum of the LE scales as,
\begin{align}
1-L_{min}(N,g) \propto g^2 N^{2/\nu}
\label{LEmin}
\end{align}
This relation indicates a critically enhanced decay of the LE that can be used to obtain the correlation length critical exponent $\nu$.
The values of $1-L_{min}$ are derived from the first minimum of biorthogonal LE from Fig.\ref{collapsefig}(a) and are plotted in Fig.\ref{collapsefig}(c) with respect to the lattice size $N$.
By fitting the data, we get the critical exponent $\nu=0.968$ which is consistent with that of the Ising transition in equilibrium.
The scaling law in Eq.(\ref{LEmin}) provides an easy way for studying the quantum criticality. 
However, it needs to know the critical value $\lambda_c$ in advance because of the condition $\delta \lambda=0$, which limits its application to unknown systems.

Motivated by the behavior of critically enhanced decay of LE and the definition of the fidelity susceptibility, we introduce a time-average rate function,
\begin{align}
\bar{r}(N,\lambda)=-\frac{1}{N}\frac{\ln(\bar{L}(N,\lambda))}{g^2},
\label{rateAver}
\end{align}
where, $\bar{L}(N,\lambda)$ is the time average LE that is defined as,
\begin{align}
\bar{L}(N,\lambda)=\frac{1}{T} \int_{0}^{T} L(N,\lambda,t)dt.
\label{LEAver}
\end{align}
Here, $L(N,\lambda,t)$ is defined in Eq.(\ref{BioLE}) with $g=\lambda_i-\lambda_f$. The time average LE that usually demands a very large time $T$ has been used 
to characterize the phase transition of Ising model in nonzero temperatures recently \cite{zhang2021quantum}. Here, will we show that a short-time average LE 
can help identifying phase transitions due to the critically enhanced decay behavior. 

To achieve it, we first find the pseudo critical point $\lambda_{N}^{\ast}$ for each lattice $N$ using Eq.(\ref{rateAver}) by varying the control parameter $\lambda$. 
The pseudo critical value $\lambda_{N}^{\ast}$ is derived from the peak of the time-average rate function as shown in Fig.\ref{ratefig}(a).
We then perform the calculations for a quench by using $\lambda_i = \lambda_{N}^{\ast}$ to $\lambda_f = \lambda_{N}^{\ast} - g$ with 
a small $g = -0.01$ to find the minimum $L_{min}(N,\lambda)$ [see Fig.\ref{ratefig}(b)].
Finally, we extrapolate the critical exponent $\nu$ from the scaling relation Eq.(\ref{LEmin}) to obtain the correlation length critical exponent [c.f. Fig.\ref{ratefig}(c)]. 
We get the critical exponent $\nu=0.993$ from fitting the data which is in agreement with that of Ising transition. Consequently, it offers a more flexible approach to study the quantum criticality. 
We note that this probe method can apply to both Hermitian and non-Hermitian many-body systems with second-order phase transitions without knowing the critical value in advance. 
More examples will be given to illustrate this approach in future research.

%%%%%%%%%%%%%%%%%%%%%%%%%%%%%%%
% Conclusion
%\section {Conclusion}
%\label{sec:Con}
\textit{Conclusion}.- In summary, we have studied the finite-size dynamical scaling of the biorthogonal LE in the one-dimensional NHTI model. 
We have shown that the LE can serve as a probe to detect biorthogonal many-body phase transitions. 
That is to say, we can probe quantum criticality of non-Hermitian many-body systems from the biorthogonal LE dynamics in experiments without knowing the exact critical values in advance. 
We note that the concept of the biorthogonal LE of the systems is general for any non-Hermitian many-body Hamiltonian with real eigenvalues. 
Therefore, it would be possible to apply the biorthogonal LE to understand the critical properties of unknown phase transitions of arbitrary non-Hermitian many-body systems 
as long as the ground states are well defined. Moreover, it would be more intriguing to know whether the scaling laws of the biorthogonal LE is able to be applicable to 
non-Hermitian many-body systems with complex eigenvalues in the future.

% Acknowledgments
%\begin{acknowledgments}
\textit{Acknowledgments}.- G.S. would like to thank W.-L. You for useful discussions and comments for the paper.
G. S. is appreciative of support from the NSFC under the Grant Nos. 11704186 and 11874220.
S. P. K  is appreciative of supported by the NSFC under the Grant Nos. 11674026, 11974053.
Numerical simulations were performed on the clusters at Nanjing University of Aeronautics and Astronautics.
%\end{acknowledgments}

%\appendix
%\section{Biorthogonal Loschmidt echo}
%\label{AppA}

\bibliographystyle{apsrev4-1}
\bibliography{refnew}
\end{document}